\documentclass[10pt,letterpaper]{article}
\usepackage{opex3}
\newcommand\beq{\begin{equation}}
\newcommand\eeq{\end{equation}}

\begin{document}

%%%%%%%%%%%%%%%%%% title page information %%%%%%%%%%%%%%%%%%
\title{Experimental evidence of cut-wire-induced enhanced transmission of transverse-electric fields through sub-wavelength slits in a thin metallic screen}

\author{Emiliano Di Gennaro,$^{1,*}$ Ilaria Gallina,$^2$ Antonello Andreone,$^1$ Giuseppe Castaldi,$^2$ and Vincenzo Galdi$^2$}

\address{$^1$CNR-SPIN and Dept. of Physics, University of Naples ``Federico II,'' I-80125 Naples, Italy\\
$^2$ Waves Group, Department of Engineering, University of Sannio, I-82100 Benevento, Italy}

\email{emiliano@na.infn.it} %% email address is required



%%%%%%%%%%%%%%%%%%%%% Created:          	03/08/2010 	%%%%%%%%%%%%%%%%%%%%%
%%%%%%%%%%%%%%%%%%%%% In final form:     	22/10/2010 	%%%%%%%%%%%%%%%%%%%%%
%%%%%%%%%%%%%%%%%%%%% To be published in Optics Express	%%%%%%%%%%%%%%%%%%%%%

\begin{abstract}
Recent numerical studies have demonstrated the possibility of achieving substantial 
enhancements in the transmission of transverse-electric-polarized electromagnetic fields
through subwavelength slits in a thin
metallic screen by placing single or paired metallic cut-wire arrays at a close distance from the screen. In this paper, we report on the first experimental evidence of such extraordinary transmission phenomena, via microwave (X/Ku-band) measurements on printed-circuit-board prototypes. Experimental results agree very well with full-wave numerical predictions, and indicate an intrinsic robustness of the enhanced transmission phenomena with respect to fabrication tolerances and experimental imperfections. 
\end{abstract}


%%%%%%%%%%%%%%%%%%%%%%%%%%%%%%%%%%%%%%%%%%%%%%%%%%%%%%%%%%%%%%%%%%
\section{Introduction}
%%%%%%%%%%%%%%%%%%%%%%%%%%%%%%%%%%%%%%%%%%%%%%%%%%%%%%%%%%%%%%%%%%
During the late 1990s, a series of seminal studies on the transmission of electromagnetic (EM) fields through arrays of sub-wavelength holes \cite{Ebbesen1,Ebbesen2} or slits \cite{Porto,Schroter,Treacy} in metallic screens demonstrated the possibility of achieving strong enhancements with respect to the very low levels predicted by the well-known Bethe's theory \cite{Bethe,Bouwkamp}. This generated 
an enormous interest in the physics and potential applications of {\em extraordinary transmission} phenomena, which have been extensively studied over the last decade. The reader is referred to \cite{Ebbesen3,deAbajo,Weiner,Vidal} for recent reviews of the main results available.  

With specific reference to the slit geometries, most of the available studies \cite{Porto,Schroter,Treacy,Ebbesen3,deAbajo,Weiner,Vidal} focus on the transverse-magnetic (TM) polarization (i.e., magnetic field parallel to the slits), for which the slits support a {\em propagating} waveguide mode at any frequency, whereas much fewer results (see, e.g., \cite{Moreno1,Crouse,Jin,Nikitin,Vidal3,Gallina}) are available for the transverse-electric (TE) case (electric field parallel to the slits), for which the slit waveguide mode is {\em evanescent} below a (nonzero) cut-off frequency. 

In particular, it was shown in some recent numerical studies \cite{Jin,Gallina} that single and paired arrays of metallic {\em cut wires} (similar to those utilized for the synthesis of magnetic or negative-index metamaterials \cite{Linden,Soukoulis2,Kivshar,Donzelli}) placed at a closed distance from the screen (with the wires centered on the slits and parallel to them) may produce substantial enhancements on the transmission of TE-polarized fields via the excitation of strongly localized resonances. Such cut-wire structures appear particularly attractive in view of the several structural parameters that can be effectively used for controlling and tuning the resonance phenomena \cite{Rhee2,Lee1,Lee2,Burokur}. However, to the best of our knowledge, the studies on these
enhanced-transmission phenomena are limited to numerical simulations, and no experimental evidence is currently available. This motivates our present study, which focuses on the experimental characterization at microwave (X/Ku-band) frequencies of single- and paired-array prototypes fabricated via standard printed-circuit-board (PCB) technology.

%############################################################
%                Figure1
%
\begin{figure}
\begin{center}
\includegraphics [width=12.cm]{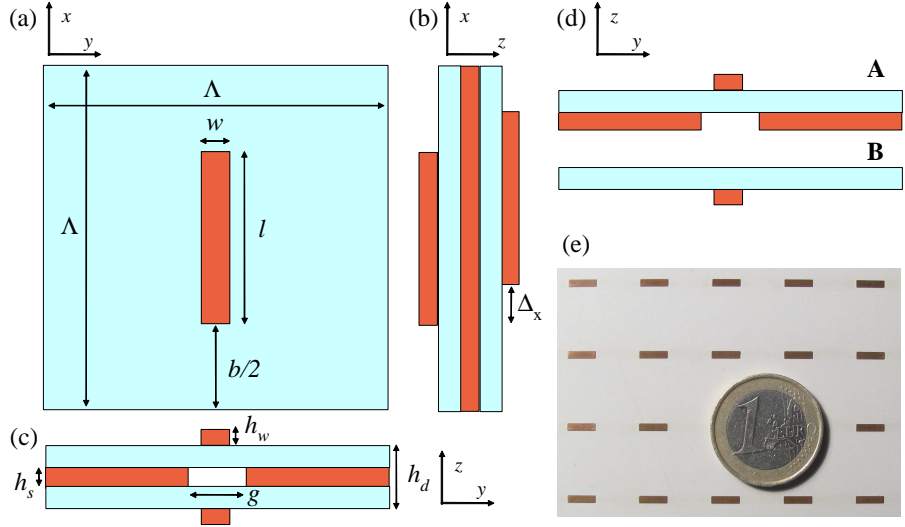}% Here is how to import EPS art
\end{center}
\caption{(a)--(c) Top and side views of the unit cell (details in the text),
with dark and bright regions representing metallic and dielectric materials,
respectively (thicknesses are exaggerated for visualization purposes). (d) Side view of the two laminates utilized for fabrication purposes. (e) Detail of the cut-wire array in the fabricated prototypes.}
\label{Figure1}
\end{figure}
%############################################################

%%%%%%%%%%%%%%%%%%%%%%%%%%%%%%%%%%%%%%%%%%%%%%%%%%%%%%%%%%%%%%%%%%
\section{Geometry, prototype fabrication, and experimental setup}
%%%%%%%%%%%%%%%%%%%%%%%%%%%%%%%%%%%%%%%%%%%%%%%%%%%%%%%%%%%%%%%%%%

Figure \ref{Figure1} illustrates the geometry of interest, which is nearly identical to that in \cite{Gallina}.
Specifically, Figs. \ref{Figure1}(a)--\ref{Figure1}(c) show the top and side views of the square unit cell of side-length $\Lambda=15$ mm, constituted by a copper screen (of electrical conductivity $5.8\times10^7$ S/m, and thickness $h_s=0.070$ mm) with an $x-$directed slit of subwavelength width $g=2$ mm covered at both sides by a dielectric substrate (of relative permittivity 3.02, loss-tangent of $0.0016$, and thickness $h_d=0.762$ mm) with $x-$directed copper wires (of length $l=5.9$ mm, width $w=1.5$ mm, and thickness $h_w=0.070$ mm) printed at each side (centered with respect to the slits) with a longitudinal gap $b=\Lambda-l=9.1$ mm, and a possible longitudinal displacement $\Delta_x$. The structure is obtained by replicating the unit cell along the $x-$ and $y-$ directions with period $\Lambda$. The above design was obtained with the aid of a full-wave commercial software package (CST Microwave Studio \cite{CST}) based on the finite-integration technique (see \cite{Gallina} for details).

For the prototype fabrication, we started with two standard Rogers 3203 laminates \cite{Rogers} composed of a polytetrafluoroethylene-based substrate cladded at both sides with electro-deposited copper foils. Via standard PCB processing, we then printed two different metal patterns, shown in Fig. \ref{Figure1}(d) as A- and B-type, respectively. For the single-array design (as in \cite{Jin}), we used only the A-type laminate, whereas the paired design (as in \cite{Gallina}) was obtained by joining the two laminates. In both cases, we used a rigid plastic framework in order to keep the laminate(s) flat and properly aligned (with a controllable displacement along the $x-$axis in the paired case).  Figure \ref{Figure1}(e) shows a detail of one of the cut-wire arrays in the fabricated prototypes.
Note that, as a consequence of the selected fabrication process, our paired design slightly differs from that in \cite{Gallina} in the slits filling (air instead of dielectric).

The fabricated prototypes were experimentally characterized by measuring the transmittance response
within the frequency region from 10--18 GHz (well below the $75$ GHz cut-off frequency of the fundamental TE mode supported by the slit), via a standard free-space measurement setup composed of a HP8720C Vector Network Analyzer and two identical transmitting/receiving horn antennas (with half-power-beamwidths of $9.4^o$ and
$10.7^o$ in the E- and H-plane, respectively, and 24 dB gain). The prototype under test was measured in air, with the slits and wires oriented horizontally, by placing a horn antenna at each side (at a distance of 1.5 m from the prototype), properly aligned (for normal incidence) and oriented so as to transmit/receive a horizontally-polarized electric field. In order to minimize the edge effects arising from the finite-size of the structures and the non-uniform illumination, the prototypes were terminated at each side by absorbing foam panels, resulting in a
nearly $300\times 150$ mm$^2$ illuminated area (comprising about $20\times 10$ unit cells). For a fixed position of the antenna and framework/absorbers, the transmittance response was finally obtained by dividing the transmitted powers measured in the presence and absence of the prototype.
Preliminary measurements were carried out to assess the sensitivity of the response to an imperfect alignment of the structure under test with the antennas. Intentional in-plane and out-of-plane misalignments of few degrees were observed not to yield significant variations in the transmission spectra.

%%%%%%%%%%%%%%%%%%%%%%%%%%%%%%%%%%%%%%%%%%%%%%%%%%%%%%%%%%%%%%%%%%
\section{Experimental and numerical results}
%%%%%%%%%%%%%%%%%%%%%%%%%%%%%%%%%%%%%%%%%%%%%%%%%%%%%%%%%%%%%%%%%%

\begin{figure}
\begin{center}
\includegraphics[width=9.5 cm]{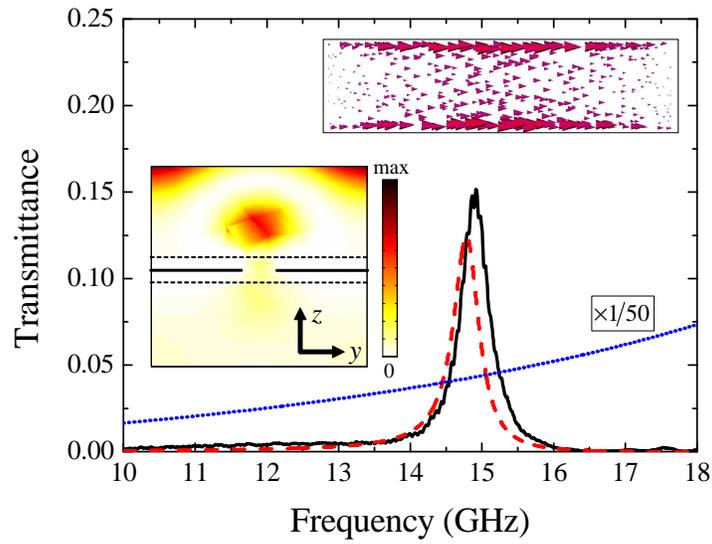}% Here is how to import EPS art
\end{center}
\caption{Measured (black solid curve) and simulated (red dashed curve) transmittance spectrum for the single-array case [A-type laminate in Fig. \ref{Figure1}(d)]. Also shown (blue dotted curve, magnified by a factor 50 for visualization purposes), as a reference, is the response in the absence of the cut-wire array. The left and right insets illustrate, at the resonance frequency of 14.78 GHz, the simulated electric field magnitude map (in the $y-z$ plane at the center of the wire gap, with the
metallic-screen and dielectric slab regions overlaid as thick-solid and dashed lines, respectively), and the corresponding surface current (real part) distribution on the wires, respectively.}
\label{Figure2}
\end{figure}

We first studied the response of the single-array prototype (as in \cite{Jin}), obtained by using only the A-type laminate in Fig. \ref{Figure1}(d). Figure \ref{Figure2} shows the measured and simulated results, from which a resonance peak is clearly visible at a frequency around 15 GHz, with a very good agreement between measurements and simulations. 
A maximum level of nearly $15\%$ of the full transmission is observed, i.e., about 170-times stronger than the response attainable in the absence of the cut-wire array (also shown as a reference).
The insets show the simulated resonant surface current distribution on the wire and the corresponding electric field intensity distribution in the $y-z$ plane (at the center of the wire gap), from which one notes the dipole-type (half-wavelength) resonance, and a strong field-localization effect responsible for the field ``squeezing'' through the slit.
%############################################################
%                Figure2
%
%############################################################
%############################################################
%                Figure3
%
\begin{figure}
\begin{center}
\includegraphics[width=9.5 cm]{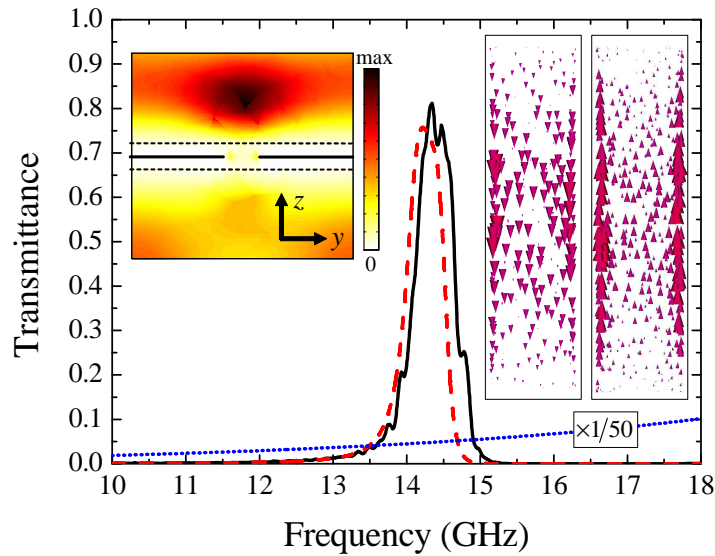}% Here is how to import EPS art
\end{center}
\caption{As in Fig. \ref{Figure2}, but for the paired-array prototype [cf. Fig. \ref{Figure1}(c)]. The insets pertain to a (simulated) resonance frequency of 14.22 GHz.}
\label{Figure3}
\end{figure}
%############################################################

We then moved on to studying the paired-array prototype obtained by joining the A- and B-type laminates [Fig. \ref{Figure1}(c)]. Figure \ref{Figure3} shows the results pertaining to the symmetric configuration (with the cut-wire longitudinally aligned, i.e., $\Delta_x=0$). Again, a resonance peak appears at a slightly lower frequency (around 14.4 GHz), with a significant increase in the intensity (now reaching nearly $80\%$ of the full transmission, i.e., about 910-times stronger than the response attainable in the absence of the cut-wire arrays) and a slightly wider bandwidth. Also in this case, measurements and simulations agree very well. 
In fact, the consistently good agreement observed in spite of the markedly different (and yet both extraordinary) transmittance levels exhibited by the two prototypes provides further confirmation of the accuracy and robustness of our measurement setup, and indicates that no artifacts are introduced by the normalization procedure involved.
From the simulated resonant surface-current and field distributions shown in the insets, we note that the current flow in the two wires is {\em opposite}, and the amplitudes are moderately different; such behavior is typical of a {\em hybrid} mode \cite{Gallina} with predominant {\em anti-symmetric} (i.e., magnetic-type) character.

%############################################################
%                Figure4
%
\begin{figure}
\begin{center}
\includegraphics[width=12 cm]{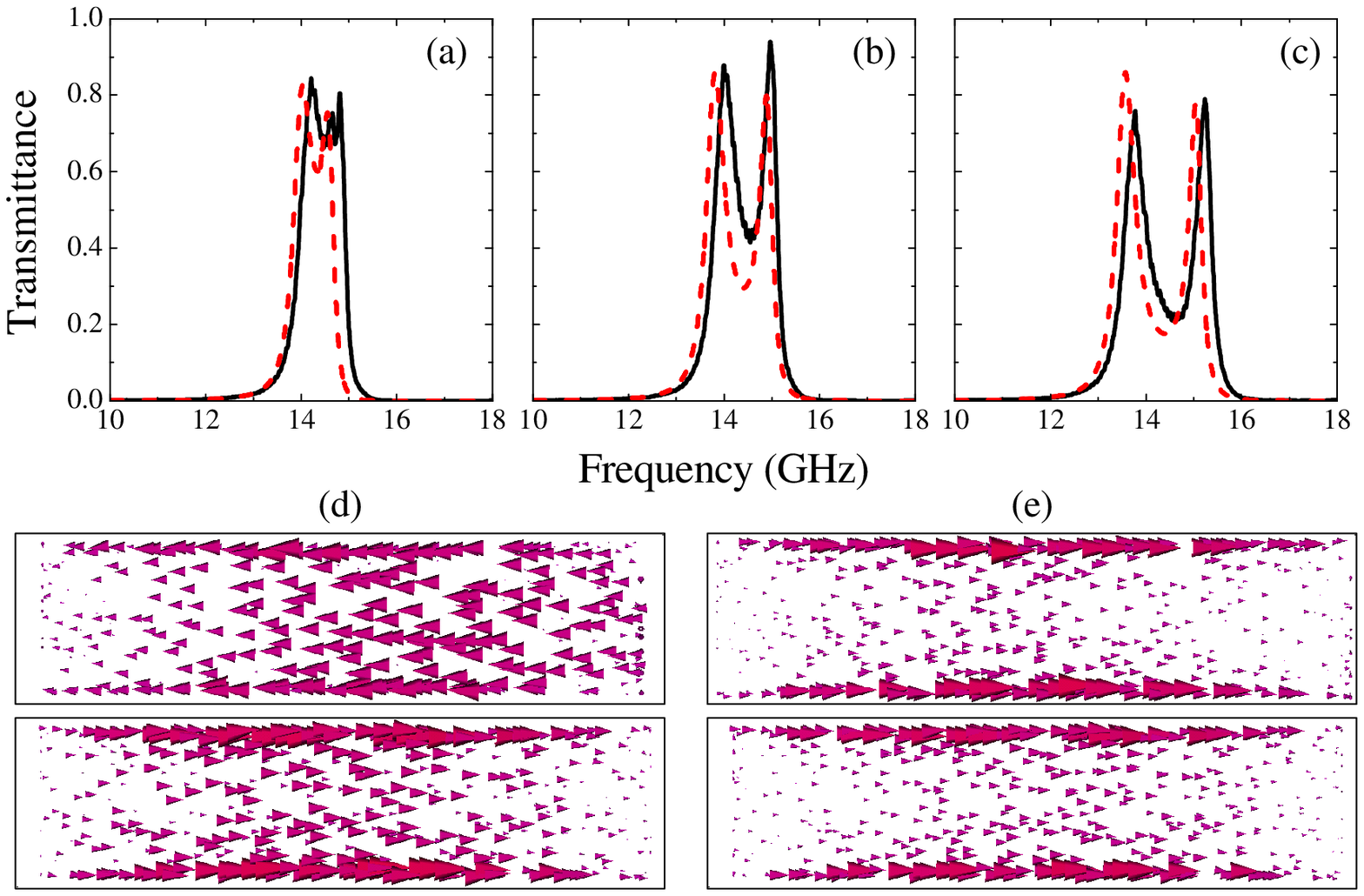}% Here is how to import EPS art
\end{center}
\caption{(a), (b), (c) As in Fig. \ref{Figure3}, but for nonzero wire longitudinal displacement,  $\Delta_x=$ 1, 2, 3 mm, respectively. (d), (e) Representative (simulated) surface current distributions on the wires for $\Delta_x=2$ mm, at the resonance frequencies of 13.82 GHz and 14.89 GHz, respectively.}
\label{Figure4}
\end{figure}

%############################################################

We highlight that the considerably higher transmittance observed in the paired-array case does not necessarily imply a superiority of such configuration as compared with the single-array case (which, as shown in \cite{Jin}, may also exhibit comparable transmittance levels). 
Our actual interest in the paired-array design is due to the possibility (illustrated in \cite{Gallina}) of exciting and tuning (jointly or separately) both {\em electric}- and {\em magnetic}-type resonances (which appear {\em merged} in the example in Fig. \ref{Figure3}). 
For instance, Fig. \ref{Figure4} illustrates some representative responses obtained by longitudinally displacing the wires by the quantity $\Delta_x$ [see Fig. \ref{Figure1}(b)] as a tuning parameter. In the transmittance responses [Figs. \ref{Figure4}(a)--\ref{Figure4}(c)], {\em two} resonance peaks are now visible, with their frequency separation increasing up to nearly 1.5 GHz (i.e., about 10\%) for higher values of $\Delta_x$, and with only mild variations of the peak transmittance levels (always around 80\%). From the representative ($\Delta_x=2$mm) surface current distributions shown in Figs. \ref{Figure4}(d) and \ref{Figure4}(e), we note that the lower frequency resonance is attributable to an anti-symmetric (i.e., magnetic-type) mode, while the higher-frequency one stems from a symmetric (i.e., electric-type) mode.
The above experimental results, again in very good agreement with the simulations, clearly indicate the practical feasibility of a {\em pass-band-type} design for cut-wire-induced enhanced transmission as proposed in \cite{Gallina}.

%%%%%%%%%%%%%%%%%%%%%%%%%%%%%%%%%%%%%%%%%%%%%%%%%%%%%%%%%%%%%%%%%%
\section{Conclusions}
%%%%%%%%%%%%%%%%%%%%%%%%%%%%%%%%%%%%%%%%%%%%%%%%%%%%%%%%%%%%%%%%%%
In conclusion, we have experimentally verified the potentials of single and paired cut-wire arrays in enhancing the transmission of TE-polarized fields through subwavelength slits in a thin metallic screen. Our results, in very good agreement with the full-wave numerical predictions, evidence a remarkable robustness of these phenomena with respect to fabrication tolerances and experimental imperfections (misalignments, edge effects, etc.), and confirm the intriguing design potentials envisaged in the previous numerical studies.

%\bibliography{EOT_bib1}

\end{document}